\documentclass{article}
\usepackage{spconf,amsmath,graphicx}

\title{Weighted Recursive Least Square Filter and Neural Network based Residual Echo Suppression for the AEC-Challenge}
\name{Ziteng Wang$^{\dagger\ddagger}$, Yueyue Na$^{\dagger}$, Zhang Liu$^{\dagger}$, Biao Tian$^{\dagger}$, Qiang Fu$^{\dagger\ddagger}$}
\address{$^\dagger$Machine Intelligence Technology, Alibaba Group \\
$^\ddagger$Beijing Sound Connect Technology}
\begin{document}
%\ninept
%
\maketitle
\begin{abstract}
This paper presents a real-time Acoustic Echo Cancellation (AEC) algorithm submitted to the AEC-Challenge. The algorithm consists of three modules: Generalized Cross-Correlation with PHAse Transform (GCC-PHAT) based time delay compensation, weighted Recursive Least Square (wRLS) based linear adaptive filtering and neural network based residual echo suppression. The wRLS filter is derived from a novel semi-blind source separation perspective.
The neural network model predicts a Phase-Sensitive Mask (PSM) based on the aligned reference and the linear filter output. The algorithm achieved a mean subjective score of 4.00 and ranked 2nd in the AEC-Challenge.
\end{abstract}
\begin{keywords}
AEC-Challenge, weighted RLS, residual echo suppression, deep neural network
\end{keywords}
\section{Introduction}
\label{sec:intro}

Acoustic Echo Cancellation (AEC) plays an essential part in full-duplex speech communication systems. The goal of AEC is no echo leakage when there is loudspeaker signal (far end) and no speech distortion when the users talk (near end). It has been a challenging problem since the earlier days of telecommunication~\cite{benesty2001advances}. A practical acoustic echo cancellation solution, e.g. the one in the WebRTC project~\cite{webrtc}, usually consists of three modules: Time Delay Compensation (TDC), linear adaptive filtering and Non-Linear Processing (NLP).

Time delay compensation is necessary, especially in real systems where microphone signal capturing and loudspeaker signal rendering are handled by different threads and the sample clocks may not be synchronized. Typical delays between the far end and near end signals range from 10 ms to 500 ms. Though in theory, the linear adaptive filter can handle any delay by having a sufficient number of filter taps. TDC could benefit the performance by avoiding over-parameterization and speeding up convergence. Time delay estimation methods include the Generalized Cross-Correlation with PHAse Transform (GCC-PHAT) algorithm~\cite{knapp1976generalized} and audio fingerprinting technology~\cite{voelcker2012robust}.

\begin{figure}[thb]
	\centering
	\centerline{\includegraphics[width=0.9\columnwidth]{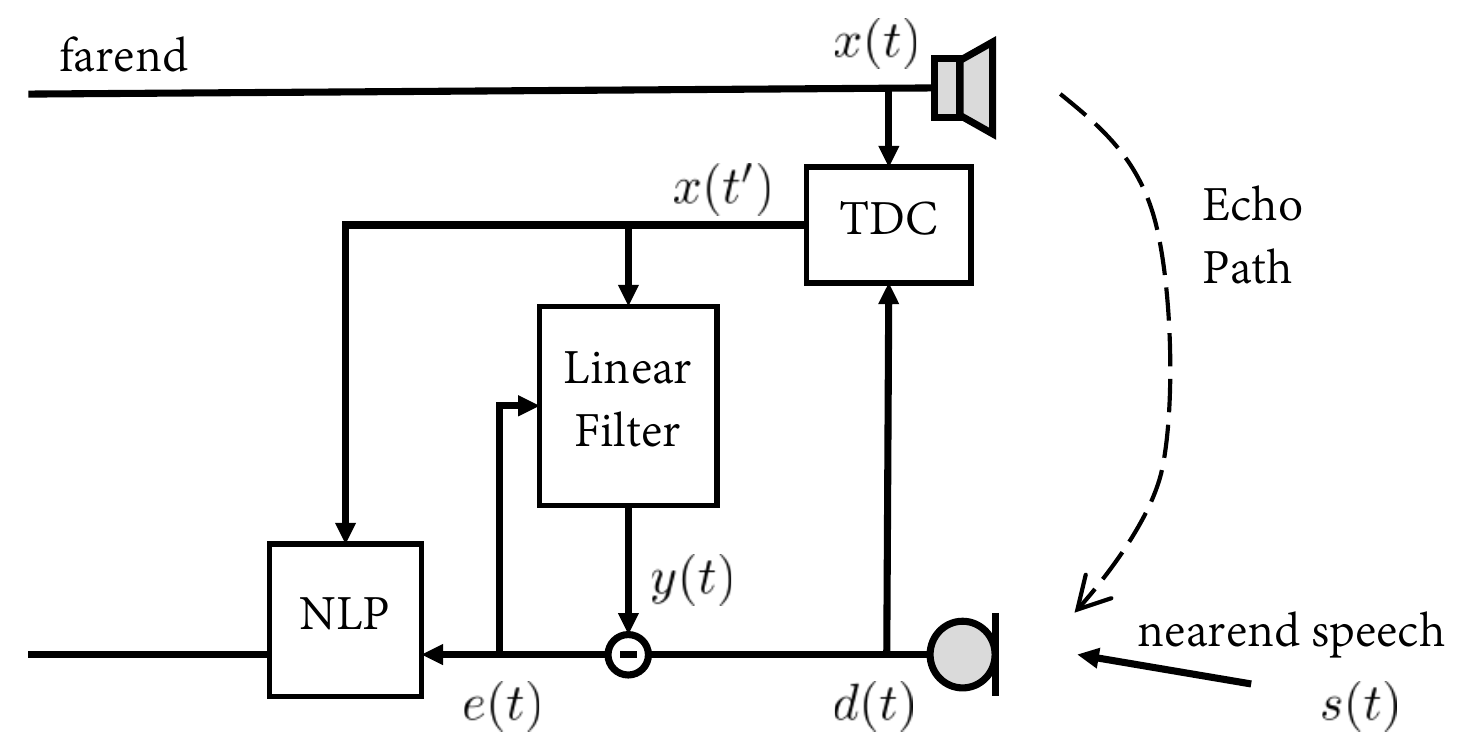}}
	\caption{A typical acoustic echo cancellation solution.}
	\label{fig1}
\end{figure}

Linear adaptive filters, such as Normalized Least Mean Square (NLMS) filters~\cite{haykin2008adaptive} and Kalman filters~\cite{wu2016robust}, can be designed either in the time domain or in the frequency domain. For the best performance possible, the filter length should be long enough to cover the whole echo path, which could be thousands of taps in the time domain. Frequency Domain Adaptive Filter (FDAF)~\cite{shynk1992frequency} are more often chosen for computational savings and better modeling statistics. 

NLP is introduced as a complement to linear filtering to suppress residual echos. The methods are generally adapted from noise reduction techniques, e.g. the multi-frame Wiener filter~\cite{huang2016multiframe}. Many recent studies also adopt deep learning methods
for residual echo suppression~\cite{carbajal2018multiple,fazel2019deep,zhang2019deep,fazel2020cad} and report reasonable objective scores on synthetic datasets. One concern is that the neural network models may degrade significantly in real applications. The AEC-Challenge~\cite{Sridar2020} is thus organized to stimulate research in this area by providing recordings from more than 2,500 real audio devices and human speakers in real environments. The evaluation is based on the average P.808 Mean Opinion Score (MOS)~\cite{naderi2020open} achieved across all different single talk and double talk scenarios.

This paper describes our submission to the AEC-Challenge, which consists of three cascading modules: GCC-PHAT for time delay compensation, weighted Recursive Least Square (wRLS) for linear filtering, and a Deep Feedforward Sequential Memory Network (Deep-FSMN)~\cite{zhang2018deep} for residual echo suppression. The wRLS filter is derived from a novel semi-blind source separation perspective and is shown to be double talk friendly. The algorithm proved its efficacy in the Challenge and it is described in the following section.

%Section 2 details the three modules. Section 3 relation to prior work. Section 4 experiments and Section 5 is conclusion.

\section{The proposed algorithm}
\label{sec3}

As in Figure~\ref{fig1}, the captured signal at time $t$ is expressed as:
\begin{equation}
	d(t) = x(t)*a(t) + s(t) + v(t)
\end{equation}
where $x(t), s(t)$ and $v(t)$ are respectively the the far end signal, the near end speech signal and the signal modeling error. $a(t)$ denotes the echo path and $*$ denotes convolution. It is assumed $v(t)=0$ in the following for simplicity. The frequency representations of $d, x, a, s$ are respectively denoted as $D, X, A, S$.

\subsection{Time Delay Compensation}

The GCC-PHAT algorithm is applied first to align the far end and near end signals. The generalized cross correlation is defined as $\Phi_{t,f} = E[X_{t,f}D^{*}_{t,f}]$ with $E[\cdot]$ denoting expectation, $f$ the frequency index and $(\cdot)^*$ the conjugate of a variable. The online implementation is given by:
\begin{equation}
\label{eq:gcc}
	\Phi_{t,f} = \alpha \Phi_{t-1,f} + (1 - \alpha) X_{t,f}D^{*}_{t,f}
\end{equation}
where $\alpha$ is a smoothing parameter. The relative delay $\tau$ is obtained by performing Inverse Fast Fourier Transform (IFFT) and finding the index of the maximum:
\begin{equation}
\label{eq:delay}
\tau = \underset{\tau}{\text{argmax}} ~\text{IFFT}(\frac{\Phi_{t,f}}{|\Phi_{t,f}|})
\end{equation}

\subsection{wRLS Filtering}
Linear filtering is performed in the frequency domain on the time-aligned signals $x(t')$ and $d(t)$. Suppose an echo path of $L$ taps, the signal model is reformulated as:
\begin{equation}
\begin{bmatrix}
D_{t,f}    \\ 
{\bf x}_{L,f}
\end{bmatrix}
= \begin{bmatrix}
1       & {\bf a}^H_{L,f}\\ 
{\bf 0} & {\bf I}
\end{bmatrix}
\begin{bmatrix}
S_{t,f}    \\ 
{\bf x}_{L,f}
\end{bmatrix}
\end{equation}
where ${\bf x}_{L,f} = [X(t',f), X(t'-1, f), ..., X(t'-L+1,f)]^T$ and ${\bf a}_{L,f} = [A(t,f), A(t-1, f), ..., A(t-L+1,f)]^T$ with $(\cdot)^T$ denoting transpose and $(\cdot)^H$ Hermitian transpose. $\bf I$ is a unitary matrix of order $L$. The near
end speech can be separated by:
\begin{equation}
\label{eq:unmix}
\begin{bmatrix}
\hat{S}_{t,f}    \\ 
{\bf x}_{L,f}
\end{bmatrix} =
{\bf B}_f
\begin{bmatrix}
D_{t,f}    \\ 
{\bf x}_{L,f}
\end{bmatrix}
\end{equation}
where $\hat{(\cdot)}$ denotes the estimate of a variable and ${\bf B}_f$ is termed the unmixing matrix. 

Equation~(\ref{eq:unmix}) clearly defines a semi-blind source separation problem. Assuming independence of $\{D_{t,f}, {\bf x}_{L,f} \}$, the unmixing matrix has this unique form as:
\begin{equation}
{\bf B}_f =
\begin{bmatrix}
1       & {\bf w}^H_{L,f}\\ 
{\bf 0} & {\bf I}
\end{bmatrix}
\end{equation}
which can be solved by the well established source source separation algorithms, such as the Independent Component Analysis (ICA) and auxiliary-function based (Aux-)ICA algorithms~\cite{ono2010auxiliary}. The Aux-ICA solution is briefly described as follows and a detailed derivation can be found in~\cite{wangsemi}. 

The Kullback-Leibler divergence is introduced as the independence measure
\begin{equation}
	J({\bf B}_f)  = \int_{S_{t,f}} \int_{{\bf x}_{L,f} } p(S_{t,f},{\bf x}_{L,f}) \log \frac{p(S_{t,f},{\bf x}_{L,f})}{q(S_{t,f},{\bf x}_{L,f})}
\end{equation}
where $p(\cdot)$ represents the source Probability Density Function (PDF) and $q(\cdot)$ the product of approximated PDF of individual sources. The loss is upper bounded by the auxiliary loss function 
\begin{equation}
\label{eq:aux}
Q({\bf B}_f, {\bf C}_f) = \sum_{i=1}^{L+1} {\bf b}_{i,f}^H {\bf C}_{i,f} {\bf b}_{i,f} + const.
\end{equation}
where ${\bf b}_{i,f}^H$ is the $i$-th row vector of ${\bf B}_f$ and
the auxiliary variable
\begin{equation}
{\bf C}_{i,f} = E[\frac{G'(r_{i,t,f})}{r_{i,t,f}}{\bf x}_{t,f}{\bf x}_{t,f}^H] 
\end{equation}
with ${\bf x}_{t,f} = [D_{t,f}, {\bf x}^T_{L,f}]^T$ and $r_{i,t,f}$ the $i$-th separated source. $G(r)$ is called the contrast function and has a relationship $G(r) = -\log p(r)$. 

Equation~(\ref{eq:aux}) can be minimized in terms of ${\bf b}_{1,f}$ as: 
\begin{align}
\nonumber {\bf b}_{1,f} &= [{\bf B}_f{\bf C}_{1,f}]^{-1}{\bf i}_1 \\
&= {\bf C}_{1,f}^{-1}{\bf i}_1 .
\end{align}
with ${\bf i}_1 = [1,0,...,0]^T$ a $L+1$ dimensional vector.
Further by applying block matrix inversion of ${\bf C}_{1,f}$, the unmixing filter coefficients are given by
\begin{align}
\label{eq:rls}
{\bf w}_{L, f} &= -{\bf R}_{L, f}^{-1}{\bf r}_{L,f}
\end{align}
where
\begin{align}
\nonumber {\bf R}_{L,f} &= E[\frac{G'(r)}{r}{\bf x}_{L,f}{\bf x}_{L,f}^H], \\
{\bf r}_{L,f} &= E[\frac{G'(r)}{r}{\bf x}_{L,f}D_{t,f}^*] .
\end{align}
The separated near end speech is obtained as:
\begin{equation}
\hat{S}_{t,f} = D_{t,f} + {\bf w}^H_f {\bf x}_{L,f}.
\end{equation}

Equation~(\ref{eq:rls}) stands for a weighted RLS filter, in which the correlation weighting factor is determined by the underlying near end source PDF. In literature, a general super-Gaussian source PDF has the form of
\begin{equation}
\label{eq:pdf}
G(D_{t,f}) = (\frac{D_{t,f}}{\eta})^{\beta}, \quad 0 < \beta \leq 2
\end{equation}
where a shape parameter of $\beta \in [0.2, 0.4]$ is suggested.

\begin{figure*}[th]
	\centering
	\includegraphics[width=15cm]{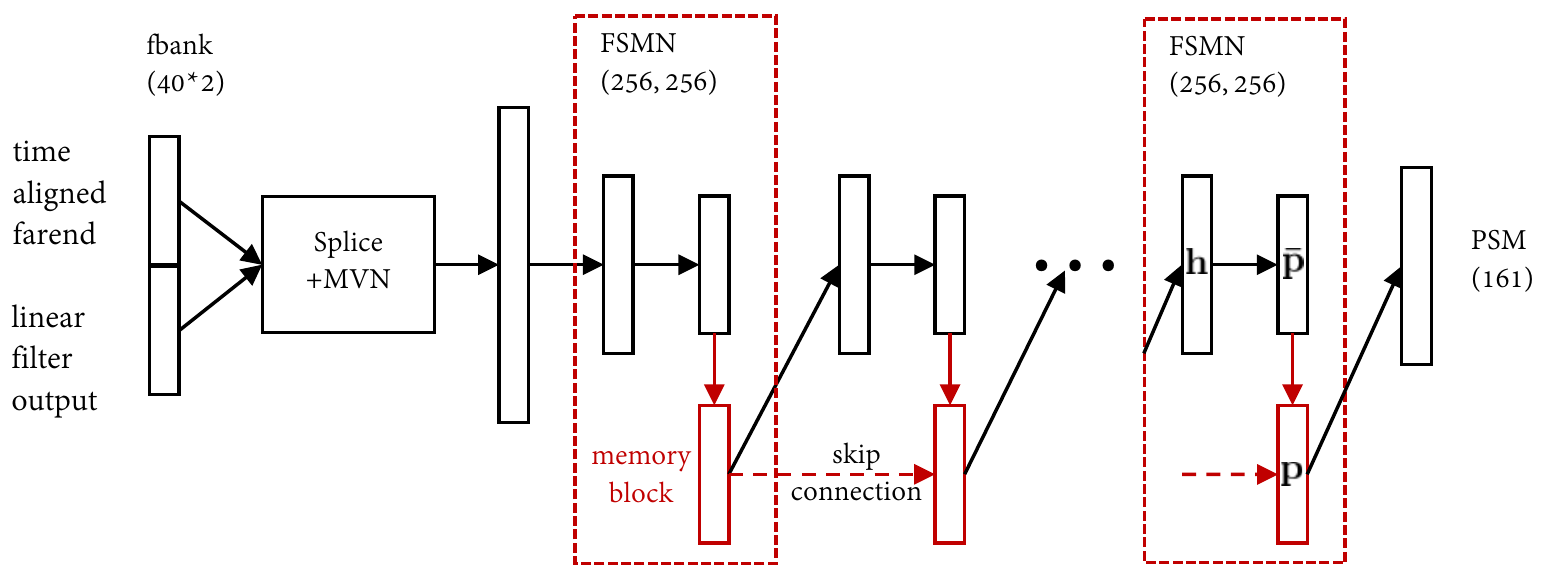}
	\caption{The Deep-FSMN model for residual echo suppression.}
	\label{fig2}
\end{figure*}

\subsection{Residual Echo Suppression}

The Deep-FSMN model for residual echo suppression is illustrated in figure~\ref{fig2}. Logarithm filter bank energies (fbank) of the time aligned far end and wRLS filter output signals are used as input to the neural network. The computation flow is given by:
\begin{align}
\label{eq:nnres}
	\nonumber {\bf f}_{in} & = [\text{fbank}(\hat{S}_t), \text{fbank}(X_{t'})] \\
	\nonumber {\bf p}^1 & = \text{ReLU}({\bf U^0}{\bf f}_{in} + {\bf v^0}) \\
	\nonumber {\bf p}^{j+1} & = \text{FSMN}({\bf p}^{j}), \quad j\in[1,2,...,J-1] \\
     {\bf f}_{out} & = \text{Sigmoid}({\bf U}^{J+1}{\bf p}^{J} +{\bf v}^{J+1})
\end{align}
where ${\bf U}^j$ and ${\bf v}^j$ are respectively the weight matrix and bias vector in the $j$-th layer. Each FSMN block has one hidden layer, one projection layer and one memory block. The realization is given by:
\begin{align}
	\nonumber {\bf h}^j_t &= \text{ReLU}({\bf U}_1^j{\bf p}^j_t + {\bf v}^j) \\
	\nonumber {\bf \bar{p}}_t &= {\bf U}_2^{j} {\bf h}^j_t \\
	{\bf p}^{j+1}_t & = {\bf p}^j_t + {\bf \bar{p}}_t + \sum_{i=0}^{N} {\bf m}^j_i\odot {\bf \bar{p}}_{t-i}
\end{align}
where ${\bf m}^j_i$ is a memory parameter weighting the history information ${\bf \bar{p}}_{t-i}$ and $\odot$ denotes element-wise multiplication. $N$ is the look-back order. Skip connections are added between the memory blocks to alleviate the gradient vanishing problem in the training phase.

The training target is a modified version of the vanilla Phase Sensitive Mask (PSM) and is clipped to the range of [0,1]
\begin{equation}
	\text{PSM} = \frac{|S_{t,f} |}{|\hat{S}_{t,f}|}\cdot  \text{Re}(\frac{S_{t,f}}{\hat{S}_{t,f}}).
\end{equation}
Though complex masks as applied in the recent DNS-Challenge~\cite{reddy2020interspeech} have potentially better performance, no significant gains are observed in our preliminary experiments.

\section{Relation to Prior work}
\label{sec4}

Addressing AEC from the source separation perspective has been investigated in~\cite{nesta2010batch,takeda2012efficient}, and ICA based solutions are discussed therein. Here, an Aux-ICA based solution is derived and results in a novel weighted RLS filter.

Exploiting deep neural networks for residual echo suppression is a trending practice in literature. Here we consider the capability of the causal Deep-FSMN architecture jointly with TDC and wRLS filter in a systematic view.

\section{Experiments}

The AEC-Challenge dataset\footnote{https://aec-challenge.azurewebsites.net/} covers the following scenarios: far end (FE) single talk (ST), with and without echo path change; near end (NE) single talk, no echo path change; double talk (DT), with and without echo path change. Both far and near end speech can be either clean or noisy. The evaluation is based on the P.808 Mean Opinion Score (MOS)~\cite{naderi2020open} on a blind test set. The top 3 results are given in Table~\ref{tab1}.

\subsection{Algorithm Details}

The wRLS adaptive filter is computed based on 20 ms frames with a hop size of 10 ms, and a 320-point discrete Fourier transform. A filter tap of $L=5$ in Equation~(4) is used, and the filter coefficients are updated as in Equation~(\ref{eq:rls}), with the correlation matrix $\bf R$ and correlation vector $\bf r$ estimated recursively using a smooth parameter of $0.8$ and a source PDF shape parameter of $\beta=0.2$ in Equation~(\ref{eq:pdf}).

The TDC part is configured to cover a relative delay of up to 500 ms, which requires a 16384-point discrete Fourier transform. To reduce the computational complexity, the estimation is updated every 250 ms by Equation~(\ref{eq:delay}) and the calculation of $\Phi_{t,f}$ in different frequencies are spread evenly in this period.

For the residual echo suppression neural network, the inference process is computed as in Equation~(\ref{eq:nnres}). The output ${\bf f}_{out}$ is point-wise multiplied with $\hat{S}_{t,f}$ for signal reconstruction.
There are $J=9$ FSMN blocks each with 256 hidden units, 256 projection units and a look-back order of $N=20$. The input feature is a spliced by one frame in the past and one frame in the future, which leads to a vector dimension of 240, and then mean and variance normalized. 

There are 1.4M trainable parameters in the model. The average time it takes to infer one frame is 0.61 ms (0.19 ms for TDC, wRLS and 0.42 ms for RES) on a Surface Laptop with Intel Core i5-8350U clocked at 1.9 GHz, based on an internal C++/SSE2 implementation.

\begin{table}[t]
	\begin{center}
		\caption{MOS across different test scenarios.}
		\label{tab1}
		\begin{tabular}{c|c|c|c|c}
			\hline
			\begin{tabular}{@{}c@{}}Team \\ Id\end{tabular}
			& \begin{tabular}{@{}c@{}}ST NE \\ MOS\end{tabular}
			& \begin{tabular}{@{}c@{}}ST FE Echo \\ DMOS\end{tabular}
			& \begin{tabular}{@{}c@{}}DT Echo \\ DMOS\end{tabular} 
			& \begin{tabular}{@{}c@{}}DT Other \\ DMOS\end{tabular} 
			\\ \hline
			21      & 3.85 & 4.19 & 4.34 & 4.07 \\ \hline
			Ours    & 3.84 & 4.19 & 4.26 & 3.71 \\ \hline
			9     & 3.76 & 4.20 & 4.30 & 3.74 \\ \hline
			Baseline & 3.79 & 3.84 & 3.84 & 3.28 \\ \hline
		\end{tabular}
	\end{center}
\end{table}

\subsection{Training Setup}

For training the neural network, the first 500 clips in the official synthetic dataset are used as the validation set and the rest 9,500 utterances are used for training. Besides, the training data is augmented as follows:

1. Randomly remix the echo and near end speech in the official synthetic dataset (19,000 utterances).

2. Select far end single talk utterances in the real dataset and randomly remix with the near end speech (28,998 utterances).

3. Use sweep signals in the real dataset to estimate the echo paths and regenerate double talk data using utterances from the LibriSpeech corpus~\cite{panayotov2015librispeech} with Signal-to-Echo Ratio (SER) uniformly distributed in [-6, 10] dB (25,540 utterances).

4. Regenerate 24,000 random room impulse responses in simulated rooms and selectively add audio effects [clipping, band-limiting, equalization, sigmoid-like transformation] to the echo signal (24,000 utterances).

The Deep-FSMN model is optimized using the Adam optimizer with a learning rate of 0.0003, under the mean squared error loss function. The model is first trained for 10 epochs on the 9,500 utterances, and then fine tuned on the augmented training set. The learning rate is decayed by 0.6 if the loss improvement is less than 0.001. The best model is selected based on the ITU-T recommendation P.862 Perceptual Evaluation of Speech Quality (PESQ) scores evaluated on the validation set.

\subsection{Analysis}

In Table~\ref{tab1}, the baseline is a recurrent neural network that takes concatenated log power spectral features
of the microphone signal and far end signal as input, and outputs a
spectral suppression mask~\cite{Sridar2020}. It performs reasonably well in the ST NE scenario, but lacks behind the top systems when echo exists. Informal listening indicates that our proposed algorithm sometimes over-suppresses the near end speech in double talk, which may explain the DT Other DMOS gap with the 1st system. 

In Table~\ref{tab2}, the proposed wRLS filter is compared with the linear filter in WebRTC-AEC3~\cite{webrtc} in terms of PESQ and Short-Time Objective Intelligibility (STOI)~\cite{falk2010non} on 500 clips of the validation set, and in terms of Echo Return Loss Enhancement (ERLE) on the ST FE in the test set. ERLE is defined as:
\begin{equation}
\text{ERLE} = 10\log_{10}\frac{E[s^2(t)]}{E[\hat{s}^2(t)]}
\end{equation} 

\begin{table}[htb]
	\begin{center}
		\caption{PESQ and STOI are evaluated on the synthetic validation set. ERLE is evaluated on the ST FE in the test set.}
		\label{tab2}
		\begin{tabular}{c|c|c||c}
			\hline
			& PESQ & STOI & ERLE (dB) \\ \hline
			Orig & 1.24 & 0.79  & -  \\ \hline
			WebRTC-AEC3 & 1.28 & 0.82 & 6.29 \\ \hline
			wRLS, $\beta=0$ & 1.41 & 0.85 & 5.58 \\ \hline
			wRLS, $\beta=0.2$ & 1.43 & 0.85 & 6.56 \\ \hline
			wRLS, $\beta=0.4$ & 1.40 & 0.85 & 5.99 \\ \hline
			wRLS, $\beta=1.0$ & 1.38 & 0.84 & 6.41\\ \hline
			\begin{tabular}{@{}c@{}}wRLS, $\beta=0.2$  \\ +Deep-FSMN\end{tabular} & 2.07 & 0.91 & 49.39 \\ \hline
		\end{tabular}
	\end{center}
\end{table}

The performance of the wRLS filter varies with different source PDF shape parameters. A value of $\beta=0.2$ is finally chosen, which outperforms AEC3 by 0.15 in PESQ, 0.03 in STOI and 0.27 dB in ERLE. The Deep-FSMN model greatly boost the overall performance, achieving a PESQ score of 2.07 and nearly complete echo reduction when echo exists.

%\vfill
%\pagebreak

\section{Conclusion}

This paper presents our submission to the AEC-Challenge. The algorithm achieves satisfactory subjective scores on real recordings by systematically combing time delay compensation, a novel wRLS linear filter and a Deep-FSMN model for residual echo suppression. The wRLS filter is derived from the semi-blind source separation reformulation of the acoustic echo cancellation problem and simplification of the Aux-ICA solution. One end-to-end neural network model that takes the raw near end mic signal and far end signal as input and outputs the near end speech is more appealing, which will be future direction of this work.

\vfill\pagebreak

\bibliographystyle{IEEEbib}
\bibliography{refs}

\end{document}